\documentclass[referee]{aa}
\usepackage{psfig}

\newcommand{\Halpha}{{\rm H}\alpha}

\psrotatefirst

\begin{document}

\thesaurus{12.03.1, 12.04.2, 10.08.1, 09.12.1}

\title{Galactic H$\alpha$ emission and the Cosmic Microwave Background}

\author{M. Marcelin\inst{1}
\and P. Amram\inst{1}
\and J.G. Bartlett\inst{2}
\and D. Valls--Gabaud\inst{2}
\and A. Blanchard\inst{2}
}

\offprints{M. Marcelin}

\institute{IGRAP, Observatoire de Marseille, 
           2 Place Le Verrier, 13248 Marseille Cedex 04, France
\and      
           Observatoire de Strasbourg, 
           11 rue de l'Universit\'e, 67000 Strasbourg, France
}

\date{Received 15 May 1998 / Accepted 24 June 1998}

\authorrunning{Marcelin et al.}
\titlerunning{H$\alpha$ emission and the CMB}

\maketitle

\begin{abstract}
	We present observations of Galactic $\Halpha$ emission 
along two declination bands where the South Pole 
cosmic microwave background experiment reports temperature
fluctuations.  The high spectral resolution of our Fabry--Perot
system allows us to separate the Galactic signal from the
much larger local sources of $\Halpha$ emission, such
as the Earth's geocorona.  For the two bands (at $\delta
= -62$\degr and $-63$\degr), we find a total mean emission of
$\sim 1$~R with variations of $\sim 0.3$~R.  The variations
are within the estimated uncertainty of our total intensity
determinations.  For an ionized gas at $T\sim 10^4$ K, this
corresponds to a maximum free--free brightness temperature of 
less than 10 $\mu$K at 30 GHz (K--band).  Thus, unless 
there is a hot gas component with $T\sim 10^6$ K, our
results imply that there is essentially no free--free contamination of
the SP91 (Schuster et al. 1993) and SP94 (Gunderson et al.
1995) data sets.

\keywords{Cosmology: --cosmic microwave background, --diffuse radiation,
Galaxy: --halo, ISM: lines and bands}

\end{abstract}

\section{Introduction}

	Cosmology has entered the era of precision cosmic
microwave background (CMB) measurements.  Since the original 
detection of temperature perturbations on large angular
scales by the COBE satellite (Smoot et al. 1992), there has been 
a myriad of new detections, resulting in a data set spanning roughly
two orders of magnitude in angular scale (Lineweaver et al. 1997;
White et al. 1994).  The extraction of cosmological information 
requires careful control and
understanding of all possible sources of signal contamination.
The current quest for high precision determination of cosmological
parameters (Jungman et al. 1996; Knox 1995)
demands a correspondingly greater understanding 
of all foregrounds.  In particular, the Galaxy, via synchrotron, 
dust and free--free emission (Bremsstrahlung), represents a source
of foreground brightness fluctuations which all experiments must recon with.  
These three contaminating emissions define a ``valley'' in
the brightness--frequency plane centered  
around 90 GHz, representing the point of smallest Galactic
contamination (Kogut et al. 1996a).  
Although, clearly, CMB efforts are concentrated in this 
``valley'', Galactic signals must nonetheless be carefully removed
to extract the purely cosmological fluctuations and to 
achieve the desired precision on cosmological parameters.

	The removal of these foregrounds is usually done in one of two ways. 
With sufficient frequency coverage and a high signal--to--noise ratio, 
a spectral analysis of the CMB data alone can in principle 
distinguish the Galactic
foregrounds from the CMB signal.  The other approach is to
use sky maps made at other frequencies as templates and to extrapolate
a given foreground emission into the CMB bands according to its 
spectral dependence.  Even when the quality of the CMB data
permits the former technique,
the second approach provides an important, {\em external} check 
on the removal procedure.  For synchrotron emission, one usually
uses the 408 MHz Haslam map (Haslam et al. 1981) and  
the 1420 MHz survey (Reich \& Reich 1986)
as a template (uncertain spatial variations of the synchrotron 
frequency index renders the procedure slightly less straightforward
than one would hope).  The IRAS all sky survey serves as a useful
template for dust emission on angular scales under $\sim 1$\degr,
and it is usually augmented with DIRBE maps on larger angular scales
(as with synchrotron emission, uncertainty in the exact slope of 
the dust emission law introduces an unfortunate complication).
 
	In this paper, we address the question of Galactic free--free
emission in relation to CMB anisotropy measurements (two recent
reviews are given by Smoot 1998 and  Bartlett \& Amram 1998).  Among the
three Galactic sources of troublesome microwave emission, free--free
emission is the most difficult to control.  This is because
the only frequency range in which it dominates over dust and synchrotron 
emission is in the CMB valley; in other words, one cannot 
extrapolate maps made at much lower or higher frequencies into 
the CMB valley to remove free--free contamination.  What
is needed is a tracer of the warm ionized interstellar medium
(WIM) responsible for free--free emission.  Given that at high
Galactic latitudes there is minimal extinction from dust, one
expects Hydrogen $\Halpha$ line emission in the excited gas to 
be a good possibility for such a tracer.

	The line emission is measured in Rayleighs (1R  
$= 10^6/4\pi\;$ photons cm$^{-2}$ s$^{-1}$ ster$^{-1}$ $= 2.41\times 10^{-7}\;$ 
erg cm$^{-2}$  s$^{-1}$ ster$^{-1}$ at $\lambda(H\alpha)=6563$ \AA)
and may be expressed in terms of the temperature and emission measure,
$EM$, of the WIM for Case B recombination:
\begin{equation}
I(H\alpha) \; \approx \; (0.36\;\mbox{R}) \; \left(\frac{EM}{\mbox{cm$^{-6}$ pc}}\right)
	\; T_4^{-0.9} \quad ,
\end{equation}
where $T_4\equiv T/10^4\;$ K; this expression is valid for temperatures
$T_4\le 2.6$ (e.g. Reynolds 1990), more accurate formulae are
given by Valls--Gabaud (1998).  
Free--free emission depends on the
same quantities (given here for pure Hydrogen and in the limit as 
$h\nu/kT\rightarrow 0$):
\begin{equation}
T_b \; \approx \; \frac{(5.43\; \mu\mbox{K})}{\nu_{10}^2 T_4^{1/2}} 
	\left(\frac{EM}{\mbox{cm$^{-6}$ pc}}\right) g_{\rm ff} \quad ,
\end{equation}
where $T_b$ is the brightness temperature, the observation frequency 
is $\nu = \nu_{10}10^{10}\;$ Hz, and $g_{\rm ff}$ is the thermally 
averaged gaunt factor, which to 20\% for $T_4\leq$ few is
\begin{equation}
g_{\rm ff} \; \sim \;  4.69 (1+0.176\ln T_4-0.118\ln \nu_{10})\quad ,
\end{equation}
(e.g. Smoot 1998). Thus, the free--free brightness associated with a given $\Halpha$ 
intensity is approximately 
\begin{equation}
\label{TbIa}
T_b \; \approx \; (15\;\mu\mbox{K}) \, g_{\rm ff} \, T_4^{0.4} \, \nu_{10}^{-2} 
	\, \left(\frac{I(H\alpha)}{\mbox{R}}\right) \quad .
\end{equation}
Valls-Gabaud (1998) discusses more accurate expressions.
	There does not, as of yet, exist a complete survey of the sky
in $\Halpha$, and the distribution of the warm ionized
medium (WIM) of our galaxy remains somewhat of a mystery.  
Local sources pose the most serious difficulties for efforts
to measure the Galactic $\Halpha$ emission.  The Earth's geocorona 
emits in $\Halpha$ with an intensity of $\sim 10\;$ R, depending 
on the season, the solar activity and the solar depression angle.  This is an order
of magnitude larger than the typical signal we expect at high
Galactic latitude.  In addition, there is an OH line
from the atmosphere at $\lambda = 6569\;$ \AA.  Fortunately,
the Earth's motion through the Galaxy displaces the 
Galactic signal relative to the local $\Halpha$ emission, and
thus the cleanest way to extract a Galactic signal is by use 
of a high--resolution spectrometer.  Reynolds has developed
this approach with a double Fabry--Perot system (Reynolds 1990)
to study the Galactic emission on degree angular scales
with pointed observations and a small--area survey 
below the Galactic Plane (Reynolds 1992; Reynolds 1980).  
This has culminated in the construction
of WHAM (Wisconsin $\Halpha$ Mapper), which is currently
surveying the entire northern sky at 1 degree resolution  
(see {\tt http://www.astro.wisc.edu/wham/}).

	Other groups have recently surveyed areas
in the north using narrow band filters (Gaustad et al. 1996;
Simonetti et al. 1996).  This technique has the advantage of much 
greater simplicity and lower cost; the inconveniences are
that one must remove the stellar contribution
by extrapolation of off--band filters and that the Geocoronal 
$\Halpha$ emission cannot be subtracted correctly.  Nevertheless,
if the geocoronal $\Halpha$ emission is stable and uniform 
across the field--of--view (survey area) during the observations,
then useful {\em upper limits} on the {\em anisotropy} of the 
Galactic signal can be obtained.  Both Gaustad et al. (1996) and
Simonetti et al. (1996) have placed limits on the possible 
contamination of CMB observations at the North Celestial Pole
and concluded that the Saskatoon (Wollack et al. 1997; Netterfield
et al. 1997) experiment is unaffected by free--free contamination.

	The situation is actually rather more complicated.
Leitch et al. (1997) have recently reported the detection of
a foreground signal around the NCP in data taken from 
the Owens Valley Radio Observatory.  The signal has
a spectral index favoring free--free emission, and it 
is well correlated with IRAS maps of the area.
Such a correlation between free--free emission and dust
emission has also been remarked by the COBE team in the DMR
data at high Galactic latitudes (Kogut et al. 1996a; 1996b).
If the foreground seen around the NCP is indeed due to Bremsstrahlung, 
then the intensity is 60 times larger than the limits implied by
the narrow band observations in $\Halpha$!  As discussed by
 Leitch et al. (1997),
this could be explained by a gas at $T\sim 10^6\;$ K, instead of 
$T\sim 10^4\;$ K.  It is interesting to note that $10^6\;$ K is 
the virial temperature of our Galactic halo. 
Although difficult to understand how, 
another possibility is that the narrow band observations
are missing something.  A further possibility is that this signal is due 
to the rotational emission of
very small spinning dust grains (Draine \& Lazarian 1998). In any case, 
present data 
are not sufficient to yield a complete understanding of the
importance of free--free contamination for CMB observations 
(Smoot 1998; Bartlett \& Amram 1998).

	In this paper, we present some of our $\Halpha$ observations 
at high galactic latitude in the Southern Hemisphere.  The 
telescope and detector system were optimized for a survey of 
the Galactic Plane in $\Halpha$ at a resolution of 9 arcsecs
(Le Coarer et al. 1992), and so it is not the most appropriate 
instrument with which to constrain the distribution of the WIM 
on CMB angular scales ($\sim 1\;$ degree).  Nevertheless, by 
summing over pixel elements in the roughly $40'\times 40'$ 
field--of--view, we have been able to reach a sensitivity of 
$\sim 1\;$ R on a scale comparable to CMB measurements.  Our
goal was to check for free--free emission in the region of sky
where Schuster et al. (1993, SP91) and Gunderson et al. (1995, SP94)
detected microwave fluctuations.\\\\\\\\\\

%
\begin{figure*}[ht]
\psfig{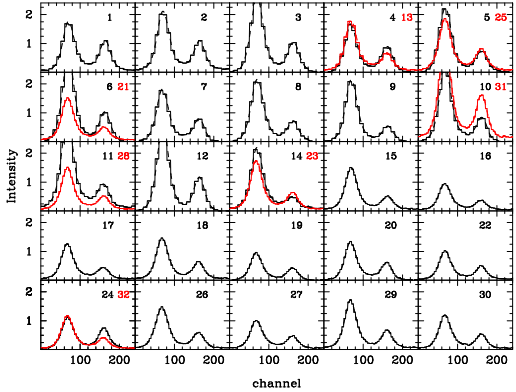}
\caption{
Summary of the observed profiles for the different fields. The reference
numbers are those of Table~2.
}
\label{fig1}
\end{figure*}

\section{Observations}

The observations were made in November 1996 with a 36 cm telescope 
in La Silla (Chile). This telescope, equipped with a
scanning Fabry-Perot interferometer, is devoted to a Survey of the 
Milky Way and Magellanic Clouds 
(Amram et al. 1991, Le Coarer et al. 1992). The field
of view is $38'\times 38'$; the spectral resolution was 
11.5~km~$\rm s^{-1}$ with the interferometer used here, and the sampling step was either 4.6~km~$\rm s^{-1}$, or 2.3~km~$\rm s^{-1}$ (i.e. 
0.10~\AA~or 0.05~\AA), depending on the scanning process adopted (24 or
48 channels over the free spectral range of 115~km~$\rm s^{-1}$, i.e. 2.5~\AA, of the Fabry-Perot interferometer). The H$\alpha$ line
observed was selected through a 8~\AA~FWHM interference filter with 70\% transmission, centered at 6563~\AA~for the observing
conditions. The lines passing through the filter are the Galactic 
H$\alpha$ line we are looking for, the geocoronal H$\alpha$
emission and the OH night sky line at 6568.78~\AA. These two parasitic 
lines are brighter than the Galactic H$\alpha$ line, the geocoronal 
line being typically twice as bright as the OH line and 10 times 
brighter than the Galactic line. The filter (3 cavities) transmission function 
is steep enough on the edges so that the two nearby, bright OH lines,
at 6553.62~\AA and 6577.28~\AA, are effectively suppressed by the filter
and may be neglected; all the more since the brighter line (6553.62~\AA)
is brought into coincidence with the OH line at 6568.78~\AA, their separation
being exactly 6 times the free spectral range of the Fabry-Perot.

In order to compare the Galactic H$\alpha$ emission fluctuations 
with the South Pole results (Schuster et al. 1993; 
Gundersen et al. 1995), we selected fields at declinations of $-63\degr$ 
(corresponding to SP91) and 
$-62\degr$ (corresponding to SP94). Our fields were
separated by 15 mn in right ascension, which is about 1\degr45\arcmin on 
the sky, thus offering a fair coverage of each
band. We observed 19 fields at $-62\degr$   
(from $\alpha$ = 23$\rm^{h}$50$^{\rm m}$ to 
$\alpha$ = 4$\rm^{h}$20$^{\rm m}$) and 6 fields
at $-63\degr$ (from $\alpha$ = 1$\rm^{h}$35$^{\rm m}$
 to $\alpha$ = 2$\rm^{h}$50$^{\rm m}$). Some of these fields were observed twice, on
different nights, to check the reproducibility of our measurements, 
and also at times  with a different spectral sampling. Table 1
summarizes the observations parameters and Table 2 gives 
the details of these observations with exposure times and 
the number of scanning steps. Figure~\ref{fig1} shows the profiles
observed in the 25 fields. 
The observing conditions were fairly
good, with some faint cirrus clouds on the nights of November 8th,
and 10th to the 13th. Only two exposures had to be cut because of heavy
clouds (number 14 and 25 in Table 2), 
and the corresponding fields were re--observed 
in good conditions with the 2$\rm^{h}$ exposure time
currently adopted. 
\\\\\\

\section{Data reduction}

\begin{figure}[h]
\psfig{figure=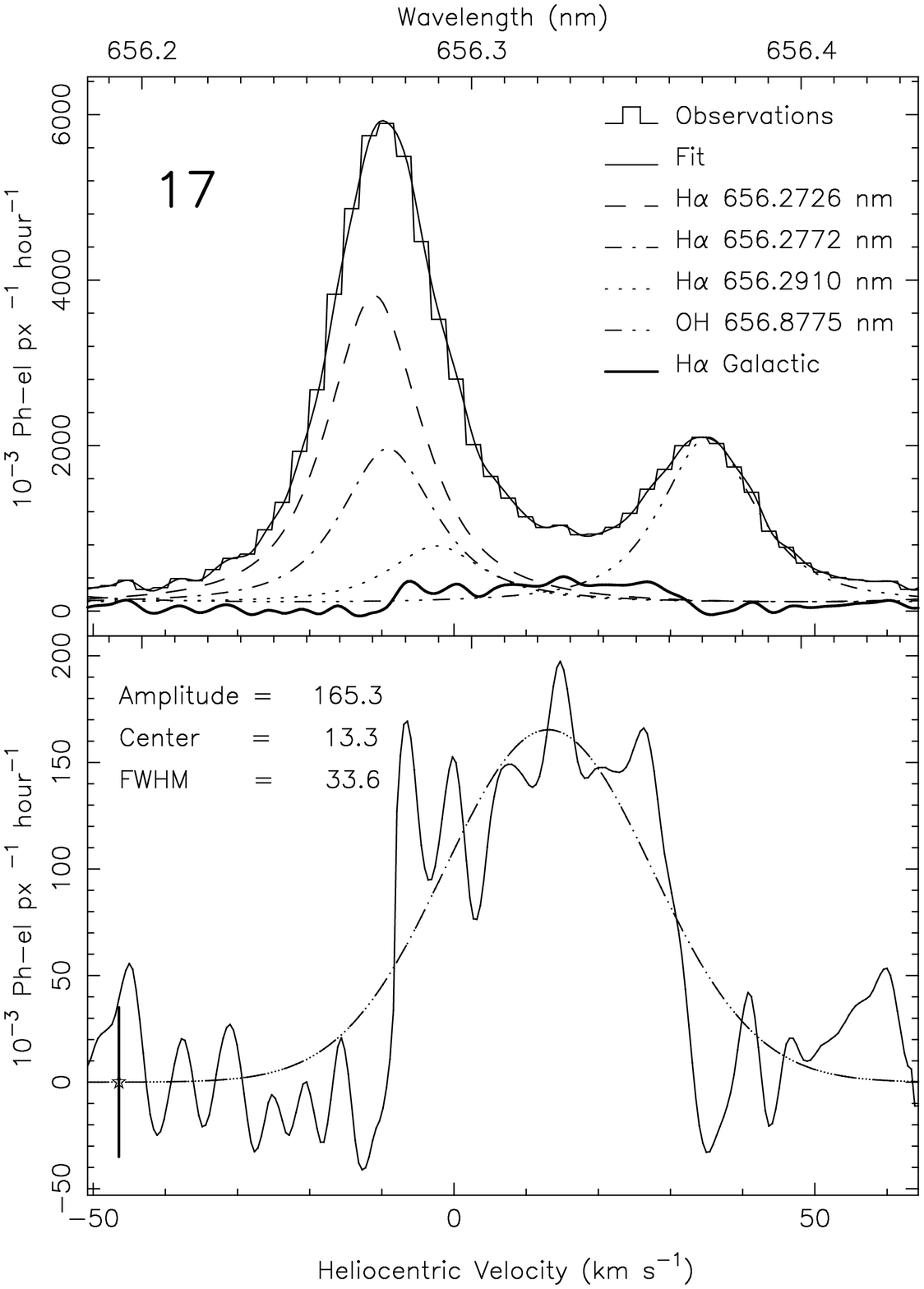,width=\columnwidth}
\caption{
Top: Example of decomposition of the observed line profile, for 
reference field 17 (see Table~2), into geocoronal 
H$\alpha$ emission (3 components) and OH night sky line. The remaining signal 
(thick line) shows the Galactic H$\alpha$ emission.
Bottom: Best fit of the Galactic H$\alpha$ emission by a gaussian 
profile. The shift with respect to the 0 km s$^{-1}$ heliocentric 
velocity is due to the motion with respect to the LSR. 
}
\label{fig2}
\end{figure}

Basically, we have to analyze a short spectrum, 2.52~\AA~wide, which is the free spectral range of our Fabry-Perot
interferometer. This means in fact that the positions of the lines are known modulo 2.52~\AA, and that there is some overlapping
of nearby lines since we select the lines to be analyzed through an 8~\AA~wide interference filter. For instance, the OH night sky line at
6568.78~\AA~appears closer to the H$\alpha$ lines 
(geocoronal and Galactic) than it actually is, with an 
apparent separation
of only 1~\AA~
(6568.78~\AA-6562.78~\AA~= 6.00~\AA~= 2 $\times$ 2.52~\AA + 0.96~\AA),
see Figure~\ref{fig2} (top panel).

The Galactic H$\alpha$ emission is slightly separated from the geocoronal emission because of our motion with respect to the
Galaxy. The motion of the Earth around the Sun and the 
motion of the Sun in the Galaxy were combined in such a manner that the
separation between the two lines remained approximately 
constant, around 0.5~\AA, along the 
bands of sky observed in November. 
As a result, the Galactic H$\alpha$ emission should appear right 
between the two parasitic night sky lines (geocoronal H$\alpha$ and
OH). Its extraction is not easy, however, since it is typically 10 
times fainter than the parasitic lines (see also Fig. 1 of 
Bartlett \& Amram 1998), whose width (FWHM
around 0.35\AA) and shape (not far from gaussian) also tend to 
bury the signal in their wings.

First of all, to improve the signal--to--noise ratio and the spectral
resolution, we selected a 30$'$ diameter disk centered on the interference rings observed in
each field. This enables us to avoid the edges of the field where the interference rings are crowded and not sufficiently sampled by
the pixel size of the image detector.
The H$\alpha$ emission profile obtained for each observed field is thus the addition of the profiles of all the pixels (about \mbox{31
000}) found within $15'$ from the center of the field.

	To analyze this profile and extract the Galactic H$\alpha$ 
emission, we must
know precisely the shape of each line to be subtracted. The OH night sky
line at 6568.78~\AA~is in fact the sum of two close components of the
same intensity, one at 6568.77~\AA~and the other at 6568.78~\AA. This fine structure can be neglected here, and the line may be considered as
a single line. More complicated is the case of the geocoronal emission line, with not less than seven fine structure transitions.
The two main components, produced by Lyman $\beta$ resonance excitation, are found at 6562.73~\AA~and 6562.78~\AA, with a 2:1
ratio (Yelle \& Roesler 1985). The resulting line center, at 6562.74~\AA, is different from that in a discharge lamp (6562.79~\AA),
where all fine structure levels are excited. However, these
two components proved insufficient when we decomposed our 
observed profiles, a residual remaining systematically at 6562.92~\AA.
This is due to cascade excitation which is particularly 
strong for the 7th component at 6562.92~\AA~(Nossal 1994). Although the
percentage of cascade contribution is not accurately known, it proved satisfactory to use the Meier model cited in Nossal's thesis,
adding a component at 6562.92~\AA~with a 1:6 ratio compared with the brightest component at 6562.73~\AA. To summarize, then, we
decomposed the H$\alpha$ geocoronal emission into three components : 6562.73~\AA, 6562.78~\AA~(with intensity ratio 1:2) and
6562.92~\AA~(with intensity ratio 1:6).

The night sky line profiles are narrow and fairly well reproduced by the instrumental profile. This profile is obtained by
scanning the emission line of a Neon lamp at 6598.95~\AA~during two hours.

%

We can thus subtract the OH line and geocoronal H$\alpha$ 
components from our observed profile since we know their positions,
the only free parameter being the intensity (remember, 
however, that the {\em relative} intensities of the three main
components we considered for the geocoronal H$\alpha$ 
line are kept fixed at 1, 1:2, 1:6). After subtraction 
of the night sky lines, a 
residual was found at the expected velocity for 
Galactic H$\alpha$ emission, that is to say around zero in V$_{LSR}$  
(radial velocity in the local standard of rest) and with the expected width, 
around 35~km~$\rm s^{-1}$, in good agreement with Reynolds' (1990)
results. Figure~\ref{fig2} shows an example of profile decomposition for one of our 
fields, together with a gaussian fit to the Galactic emission below. 
The width 
of the gaussian was left as a free parameter and adjusted automatically for 
the best fit. This width was typically 
found to lie between 25 an 50~km~$\rm s^{-1}$ (see Table 2). 
For 7 out of our 32 
observations, the signal--to--noise ratio was too faint for a good fit,
and we imposed the 
average width ($\sim$35~km~$\rm s^{-1}$).

We scanned many fields with 48 steps, instead of the usual 
24 steps, in order to check the interest of oversampling. Because
of the {\it Finesse} of our Fabry-Perot interferometer (about 10 at H$\alpha$),
the usual sampling criteria indicate that 24 scanning steps
are sufficient to obtain profiles with the best achievable resolution. 
However, oversampling is sometimes necessary, especially when decomposing
a profile into several close components. Indeed,
we found no significant difference between the observations
at 48 scanning steps and those at 24 steps, although the profiles drawn 
with 48 steps are smoother, precisely
because of the better sampling.

The calibration in intensity was made by observing through the same instrument (although through a slightly redshifted H$\alpha$
interference filter) the H{\sc ii} region N11E in the Large Magellanic Cloud, for which an absolute calibration has been performed by Caplan \&
Deharveng (1985).

Let us note that we also used a rough method which produced nearly the same intensity variations. This method consists of assuming that
the night sky lines are symmetric; then, taking into account the left wing 
of the geocoronal H$\alpha$ line, which is contaminated by neither the
Galactic emission nor the OH, one can infer that the right wing 
is its mirror image. Similarly, one considers the right wing of 
the OH night sky line,
which is not contaminated by either Galactic emission or 
geocoronal H$\alpha$, and assumes that the left wing is its 
mirror image. The Galactic
H$\alpha$ emission is then deduced from the subtraction of these two 
symmetric lines.  The results are not significantly different from the
results obtained with our more sophisticated method.

\section{Results}

We find an intensity of Galactic emission in the observed bands, at 
$-62\degr$ and $-63\degr$, varying between 0.2 R and 1.4 R
(see Table 2), in good agreement 
with intensity values measured by Reynolds (1990) in the northern hemisphere 
far from the Galactic plane.

\begin{figure}[ht]
\psfig{figure=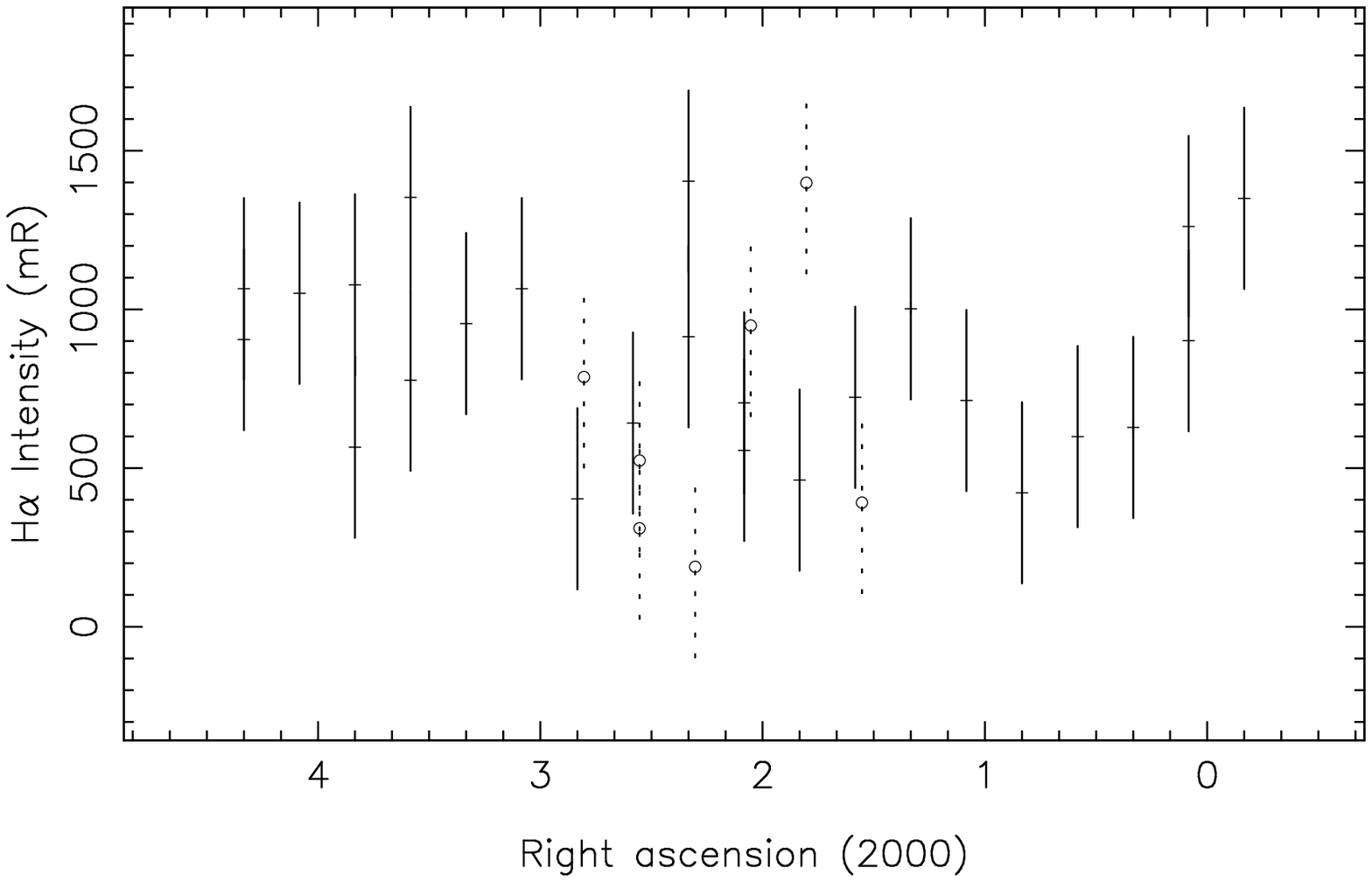,width=\columnwidth}
\caption{
H$\alpha$ intensities measured at declination $-62\degr$ (full lines)
and at  declination $-63\degr$ (dotted lines).
}
\label{fig3}
\end{figure}

Figure~\ref{fig3} shows the measured intensity of Galactic H$\alpha$ at 
declination $\delta = -62$\degr and $\delta =
-63$\degr. The error bars are the average rms difference between
the signal and the fitted gaussian, found to be 0.35 Rayleigh. 
We note that Reynolds (1990) quotes comparable uncertainties,
$\sim 0.4$ Rayleigh. 

%

To check the reproducibility of our observations, we observed seven fields
twice. The differences in intensity we found for these fields vary by
40\% in average, with just 0.4 Rayleigh as an average value, close to the
uncertainty mentioned above. 
For the seven fields observed twice, the corresponding differences in
intensity are (in \% and in increasing order): 15, 21, 28, 43, 47, 52 and 
79. The 43\% difference may be easily explained by meteorologic effects,
the lower intensity value having been obtained in bad conditions (the exposure
 had to be cut at 2640s because of clouds). The strong 79\%
difference may be explained, at least partially, by a $10'$ drift of the field 
due to the loss of the guide star at mid--exposure. However, the 
remaining 47 and 52\% differences are abnormally large and cannot be explained 
by observing conditions.

Figure~\ref{fig3} suggests that the overall variations are
small and that the galactic H$\alpha$ emission varies 
smoothly along the two bands of sky observed.

\section{Discussion and conclusions}

	The goal of these observations was to constrain and
quantify the possible Galactic free--free contamination of the 
SP91 and SP94 CMB results.  The former consists of data taken 
along a strip at $-63\degr$ declination over a narrow range 
of frequencies centered on 30 GHz (K band), and which  
show a falling signal more characteristic of
free--free emission than of the CMB.  The SP94 scan being adjacent
in declination, we also chose to observe along this band 
at $\delta=-62\degr$, although these data were taken in both K and Q 
($38-40$ GHz) bands and show fluctuations consistent with a thermal spectrum 
(CMB).     

\begin{figure}[h]
\psfig{figure=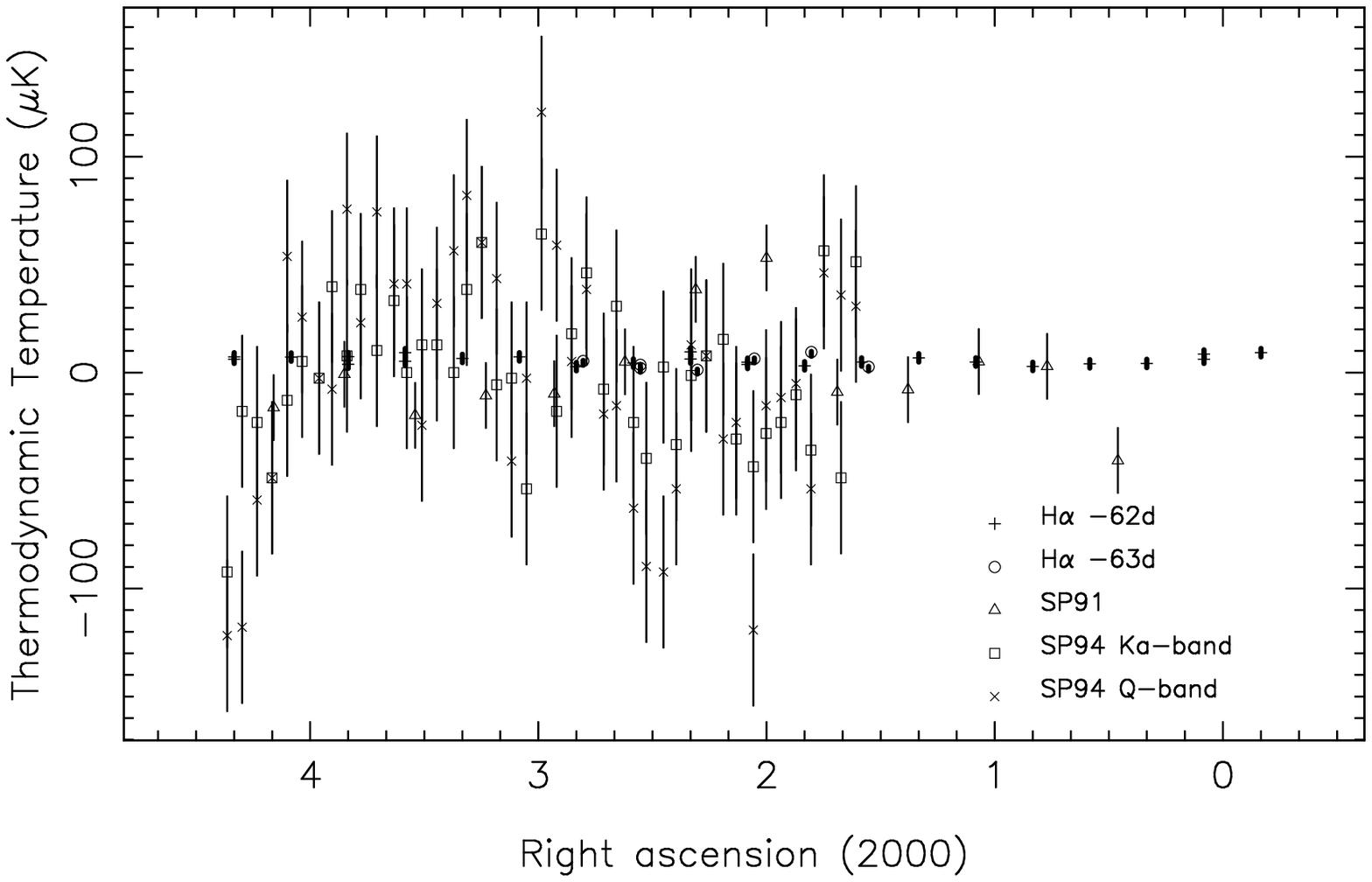,width=\columnwidth}
\caption{
Comparison of the SP results with the free-free emission signal
deduced from our H$\alpha$ observations 
}
\label{fig4}
\end{figure}

  	The absolute intensity of the total $\Halpha$ emission
over the observed bands is quite low (mean $= 0.85$ R and maximum
$< 1.5$ R), with variations equal to our estimated uncertainty
of $\sim 0.35$ R.  As 
mentioned, this is consistent with previous work using both interferometers
and narrow band filters.  If we assume the WIM producing our 
signal is indeed at $T\sim 10^4$, then Eq.~(\ref{TbIa}) indicates 
that it contributes at most $\sim 10 \mu$K to the SP results in the K band.
For comparison, the two largest fluctuations in SP91, those dominating the 
signal attain $\sim 50 \mu$K, while the {\em rms} level 
seen in SP94 is $\sim 40 \mu$K.  A comparison of our
$\Halpha$ results and the SP data sets is given in Figure~\ref{fig4}, again
using Eq.~(\ref{TbIa}).
It must be remembered that the SP points in this
figure are really differences
between fields adjacent on the sky, while the $\Halpha$ points
represent absolute intensities at the given positions.
Our results imply that, at a temperature of a few 10$^4$ K, the corresponding
free-free emission is smaller than about 10 $\mu$K at 30 GHz (K band), and 
hence does not significantly contaminate the SP experiments.  It is worth
mentioning, however, two possibilities which could lead to more important
contamination of CMB signals despite low measured H$\alpha$ intensities.
Firstly,  
the above discussion assumes that the WIM does not have a higher
temperature component.  We note that even $\sim 0.5$ R variations,
allowed at the $\sim 2\sigma$--level, from a gas at $\sim 10^6$ K 
would produce fluctuations in K--band comparable to those 
observed in the SP data sets; the spectral information from
the SP94 observations makes this seem unlikely, however.
Secondly, it may well be that an important source of foreground to the
contamination arises not from
Bremsstrahlung, but from the rotational emission of
very small spinning dust grains (Draine \& Lazarian 1998). In this
case, a correlation between the 30 GHz emission and the diffuse
12 $\mu$m emission is expected. It remains to be seen whether these
two effects will indeed be significant.

%

{\bf Acknowledgements}\\

We thank the participants to the meeting organised in November 1995
by D. Valls--Gabaud and J.P. Sivan at the Observatoire de
Haute Provence, for stimulating discussions.

%
\begin{table*}[tp]
\caption{Observations parameters}
\begin{flushleft}
\begin{tabular}[t]{lll}
\hline
\noalign{\medskip}
Observations & Telescope & Marseille 36 cm \\
             & Location & La Silla \\
             & Equipment & CIGALE  Cassegrain focus \\
             & Date & Nov 1996 \\
Interference Filter &   Central wavelength at 10$\degr$C& 6566 \AA \\
                    &   Transmission & 0.68 \\
                    &   FWHM & 8 \AA \\
Calibration & H$\alpha$ Comparison light & $\lambda$ 6562.78 \AA \\
Perot--Fabry & Interference order & 2604 @ 6562.78 \AA \\
                 & Free spectral range & 115 km s$^{-1}$ \\
                 & Finesse at H$\alpha$ & 10 \\
Spectral Sampling& 24 Scanning steps & 0.105 \AA\ (4.6  km s$^{-1}$) \\
                 & 48 Scanning steps & 0.052 \AA\ (2.3  km s$^{-1}$) \\
Spatial Sampling & Total field & $38'\times 38'$ (256$\times $256 px$^2$) \\
         & Useful circular field & $\phi~30'$  \\
         & Pixel size & $9.1''$  \\
Typical exposures times & Total exposure & 2 hours \\
                & Elementary scanning exposure time & 5 s per channel \\
                & Total exposure time per channel & 300 s\\
\noalign{\medskip}
\hline

\end{tabular}
\end{flushleft}
\end{table*}

%
\begin{table*}[bp]
\caption{Observed Fields}
\begin{flushleft}
 \begin{tabular}{rrccrrrrr}
\noalign{\medskip}
\hline
\multicolumn{1}{c}{Reference} &
\multicolumn{2}{c}{Coordinates (2000)} & 
\multicolumn{1}{c}{Date} &
\multicolumn{1}{c}{$t_{exp}$} & 
\multicolumn{1}{c}{$N_{\rm scans}$} & 
\multicolumn{1}{c}{I(H$\alpha$)} & 
\multicolumn{1}{c}{FWHM} &     
\multicolumn{1}{c}{$V_{\rm LSR}$} \\ 
\multicolumn{1}{c}{number}   & 
\multicolumn{1}{c}{$\alpha$} & 
\multicolumn{1}{c}{$\delta$} &
   & 
\multicolumn{1}{c}{s}   & 
   &
\multicolumn{1}{c}{mR} &
\multicolumn{1}{c}{km s$^{-1}$} & 
\multicolumn{1}{c}{km s$^{-1}$} \\
\hline
3  & 23$\rm^{h}$50$\rm^{m}$& $-$62\degr & Nov ~5 1996 & 7200 & 24 & 1350 & 40 &  $-$1 \\
6  & 0$\rm^{h}$05$\rm^{m}$ & $-$62\degr & Nov ~6 1996 & 7200 & 24 & 1261 & 48 &  $-$5 \\
21 &                       & $-$62\degr & Nov 11 1996 & 7200 & 48 &  901 & 31 &  +1 \\
9  & 0$\rm^{h}$20$\rm^{m}$ & $-$62\degr & Nov ~7 1996 & 7200 & 24 &  628 & 72 &  $-$2 \\
12 & 0$\rm^{h}$35$\rm^{m}$ & $-$62\degr & Nov ~8 1996 & 7200 & 24 &  599 & 34 &  $-$8 \\
15 & 0$\rm^{h}$50$\rm^{m}$ & $-$62\degr & Nov ~9 1996 & 7200 & 48 &  422 & 27 &  +4 \\
18 & 1$\rm^{h}$05$\rm^{m}$ & $-$62\degr & Nov 10 1996 & 7200 & 48 &  713 & 34 &  $-$2 \\
7  & 1$\rm^{h}$20$\rm^{m}$ & $-$62\degr & Nov ~6 1996 & 7200 & 24 & 1002 & 35 &  +4 \\
1  & 1$\rm^{h}$35$\rm^{m}$ & $-$62\degr & Nov ~4 1996 & 7200 & 24 &  723 & 34 & +10 \\
2  & 1$\rm^{h}$50$\rm^{m}$ & $-$62\degr & Nov ~4 1996 & 7200 & 24 &  462 & 38 &  +7 \\
4  & 2$\rm^{h}$05$\rm^{m}$ & $-$62\degr & Nov ~5 1996 & 7200 & 24 &  705 & 35 & +12 \\
13 &                       & $-$62\degr & Nov ~8 1996 & 7200 & 24 &  556 & 34 & +10 \\
10 & 2$\rm^{h}$20$\rm^{m}$ & $-$62\degr & Nov ~7 1996 & 7200 & 24 &  914 & 36 &  +8 \\
31 &                       & $-$62\degr & Nov 15 1996 & 7680 & 48 & 1404 & 46 &  +8 \\
16 & 2$\rm^{h}$35$\rm^{m}$ & $-$62\degr & Nov ~9 1996 & 7200 & 48 &  642 & 30 &  +7 \\
19 & 2$\rm^{h}$50$\rm^{m}$ & $-$62\degr & Nov 10 1996 & 7200 & 48 &  403 & 26 & +13 \\
20 & 3$\rm^{h}$05$\rm^{m}$ & $-$62\degr & Nov 10 1996 & 7200 & 48 & 1065 & 34 & +13 \\
17 & 3$\rm^{h}$20$\rm^{m}$ & $-$62\degr & Nov ~9 1996 & 7200 & 48 &  955 & 30 & +13 \\
14 & 3$\rm^{h}$35$\rm^{m}$ & $-$62\degr & Nov ~8 1996 & 2640 & 24 &  777 & 31 & +14 \\
23 &                       & $-$62\degr & Nov 11 1996 & 7200 & 48 & 1353 & 35 & +11 \\
11 & 3$\rm^{h}$50$\rm^{m}$ & $-$62\degr & Nov ~7 1996 & 7200 & 24 & 1077 & 35 & +13 \\
28 &                       & $-$62\degr & Nov 13 1996 & 7200 & 48 &  566 & 34 & +15 \\
8  & 4$\rm^{h}$05$\rm^{m}$ & $-$62\degr & Nov ~6 1996 & 7200 & 24 & 1051 & 29 & +10 \\
5  & 4$\rm^{h}$20$\rm^{m}$ & $-$62\degr & Nov ~5 1996 & 7200 & 24 & 1065 & 46 & +14 \\
25 &                       & $-$62\degr & Nov 12 1996 & 6960 & 48 &  905 & 28 & +11 \\
22 & 1$\rm^{h}$35$\rm^{m}$ & $-$63\degr & Nov 11 1996 & 7200 & 48 &  391 & 28 &  +6 \\
26 & 1$\rm^{h}$50$\rm^{m}$ & $-$63\degr & Nov 13 1996 & 7200 & 48 & 1399 & 38 &   0 \\
27 & 2$\rm^{h}$05$\rm^{m}$ & $-$63\degr & Nov 13 1996 & 7200 & 48 &  949 & 29 &  +7 \\
29 & 2$\rm^{h}$20$\rm^{m}$ & $-$63\degr & Nov 14 1996 & 7200 & 48 &  189 & 34 &  $-$6 \\
24 & 2$\rm^{h}$35$\rm^{m}$ & $-$63\degr & Nov 12 1996 & 7200 & 48 &  310 & 26 &  +1 \\
32 &                       & $-$63\degr & Nov 15 1996 & 7200 & 48 &  524 & 34 & +12 \\
30 & 2$\rm^{h}$50$\rm^{m}$ & $-$63\degr & Nov 14 1996 & 7200 & 48 &  787 & 28 &  +3 \\
\hline
\end{tabular}
\end{flushleft}

\end{table*}

\end{document}